# Superconductivity in the bcc-type High-entropy Alloy TiHfNbTaMo

*Lingyong Zeng[1#], Jie Zhan[2,3 #], Mebrouka Boubeche[4], Kuan Li[1], Longfu Li[1], Peifeng Yu[1], Kangwang Wang[1], Chao Zhang[1], Kui Jin[4,5,6], Yan Sun[2, 3*], Huixia Luo[1,*]*

[1]School of Materials Science and Engineering, State Key Laboratory of Optoelectronic Materials and Technologies, Key Lab of Polymer Composite & Functional Materials, Guangdong Provincial Key Laboratory of Magnetoelectric Physics and Devices, Sun Yat-Sen University, No. 135, Xingang Xi Road, Guangzhou, 510275, P. R. China

[2]Shenyang National Laboratory for Materials Science, Institute of Metal Research, Chinese Academy of Sciences, Shenyang, 110016, China;

[3]School of Materials Science and Engineering, University of Science and Technology of China, Shenyang 110016, China

[4]Songshan Lake Materials Laboratory, Building A1, University Innovation Town, Dongguan City, Guang Dong Province, 523808 China

[5]Beijing National Laboratory for Condensed Matter Physics, Institute of Physics, Chinese Academy of Sciences, Beijing, China

[6]Key Laboratory of Vacuum Physics, School of Physical Sciences, University of Chinese Academy of Sciences, Beijing, China

[#]L. Zeng and J. Zhan contributed equally to this work
E-mail: luohx7@mail.sysu.edu.cn; or sunyan@imr.ac.cn

**Abstract:** X-ray powder diffraction, electrical resistivity, magnetization, and thermodynamic measurements were conducted to investigate the structure and superconducting properties of TiHfNbTaMo, a novel high-entropy alloy possessing a valence electron count (VEC) of 4.8. The TiHfNbTaMo HEA was discovered to have a body-centered cubic structure (space group $Im\bar{3}m$) with lattice parameters $a$ = 3.445(1) Å and a microscopically homogeneous distribution of the constituent elements. This material shows type-II superconductivity with $T_c$ = 3.42 K, lower critical field $\mu_0H_{c1}(0)$ = 22.8 mT, and upper critical field $\mu_0H_{c2}(0)$ = 3.95 T. Low-temperature specific heat measurements show that the alloy is a conventional s-wave type with a moderately coupled superconductor. First-principles calculations show that the density of states (DOS) of the TiHfNbTaMo alloy is dominated by hybrid $d$ orbitals of these five metal elements. Additionally, the TiHfNbTaMo HEA exhibits three van Hove singularities. Furthermore, the VEC and the composition of the elements (especially the Nb elemental content) affect the $T_c$ of the bcc-type HEA.

# 1. Introduction

High-entropy alloys (HEAs) typically contain five or more metal elements mixed in equimolar or near-equimolar ratios.[1-3] Due to their exceptional mechanical and physical properties, HEAs, a novel category of materials, have attracted much attention in the past two decades.[4,5] Illustrative examples include superior hardness, high tensile strength, and excellent thermal and structural stability.[6-10] The novel properties of HEAs are thought to be caused by their large atomic disorder.[3,11] The field of solid-state physics has revealed one of the most captivating occurrences in HEAs: superconductivity, which has generated significant fascination regarding their pairing mechanism. In general, high levels of disorder in crystalline superconductors limit the formation of Cooper pairs, partly due to a decrease in state density or an increase in effective Coulomb repulsion between paired electrons.[12,13] It has been reported that HEA superconductors exist in bcc-,[2,3,14-22] α-Mn-,[23] β-Mn,[24] hcp-,[25-27] CsCl-,[28,29] A15-type[30-32] and σ-phase[33,34] structures. Additionally, high-entropy superconductors' superconducting transition temperature (Tc) are robust against physical pressure and show a certain correlation with valence electron count (VEC).[17,35,36] It is also worth noting that the strongly coupled behavior and high Kadowaki-Woods ratio in $Ti_{1/6}Zr_{1/6}Hf_{1/6}Nb_{2/6}Ta_{1/6}$ HEA superconductor, which could be considered a strongly correlated system.[18]

Among elements in nature, Nb has the highest $T_c$ (9.2 K). And niobium-titanium alloy is well-known, with a $T_c$ of about 10 K at zero magnetic fields.[37,38] The composition of elements is a crucial factor to consider when designing a bcc HEA superconductor. Based on the VEC requirement of the Matthias rule, bcc HEA superconductors primarily comprise Ti, Zr, Hf, V, Nb, and Ta, with a VEC of either 4 or 5.[39,40] And the $T_c$ exhibits a broad peak around 4.6 VEC. The examination of bcc HEA superconductors with VEC values surpassing 4.6 is of significant interest, as this particular range remains relatively unexplored. The systematic exploration of bcc HEA superconductors with VECs exceeding 4.6 is still nascent. Nonetheless, a scarce bcc HEA superconductor with a VEC of 4.8 or higher has solely been documented in NbTaTiZrFe, $Nb_{20}Re_{20}Zr_{20}Hf_{20}Ti_{20}$, and $(TiZrNbTa)_{1-x}W_x$.[19,20,41] In combination with late-transition metals, predominantly Cr, Mo, and W, achieving a VEC equal to 4.8 is possible. Furthermore, the bcc-type HEA superconductors with a VEC larger than or equal to 4.8 will provide more material platforms for studying the relationship between VEC and superconducting properties. Therefore, to achieve a VEC equal to 4.8 in a bcc HEA, we use the Ti, Hf, Nb, Ta, and Mo elements. The most widely equiatomic is adopted, that is, TiHfNbTaMo HEA.

This paper describes the physical properties of the equiatomic HEA compound TiHfNbTaMo, which has a VEC of 4.8 in a body-centered cubic (bcc) structure. The powder x-ray diffraction (PXRD) characterization agreed with Rietveld fitting that TiHfNbTaMo crystallizes as a bcc structure with space group $Im\bar{3}m$. TiHfNbTaMo HEA shows type-II superconductivity with $T_c$ = 3.42 K, upper critical field $\mu_0H_{c2}(0)$ = 3.95 T, and lower critical field $\mu_0H_{c1}(0)$ = 22.8 mT. The specific heat measurements indicate that the TiHfNbTaMo alloy is a BCS fully gapped s-wave superconductor.

## 2. Experimental Section

We synthesized TiHfNbTaMo HEA in an argon atmosphere by a conventional arc-melting method. Elemental titanium (99%, Alfa Aesar), hafnium (99.9%, Aladdin), niobium (99.8%, Alfa Aesar), tantalum (99.9%, Alfa Aesar), and molybdenum (99.8%, Aladdin) for TiHfNbTaMo were utilized as raw materials. A total of 0.25 g of the elemental powders in a molar ratio of 1:1:1:1:1 was pressed into a piece, then arc-melted under 0.5 atm of argon gas. Multiple flips and remelts were performed to ensure homogeneity during synthesis. Multiple flips and remelts were performed to ensure homogeneity during synthesis. PXRD was utilized to ascertain the crystal structure and phase purity. As part of the experiment, the XRD data were collected using the Rigaku MiniFlex (Cu Kα1 radiation) from 10 º to 100 º with a constant scanning speed of 1 º/min. The arc-melted samples were ground into powder for XRD analysis. To confirm the actual element ratio of the TiHfNbTaMo sample, the scanning electron microscope (SEM), back-scattered electron micrograph (BSEM), and energy dispersive X-ray spectroscopy (EDX) with an electron acceleration voltage of 20 KV were performed. The resistivity, magnetization, and heat capacity measurements were conducted using a physical property measurement system (PPMS, Quantum Design). The resistance measurements were performed with a four-probe method. The magnetization and heat capacity measurements use small pieces of sample.

The electronic structure was calculated by using the code of Vienna Ab initio Simulation Package (VASP)[48] based on density functional theory with projected augmented wave potential. The exchange and correlation energies were considered in the generalized gradient approximation (GGA), with Perdew–Burke–Ernzerhof parametrization scheme.[49] The energy cut off of plane wave basis was 500 eV. Considering the similar shell electronic states of the five elements used in this material, we used the virtual crystal approximation[50] with mixed potential and experimentally measured lattice constants.

## 3. Results and Discussion

Figure 1(a) displays the PXRD pattern for the TiHfNbTaMo HEA. Data from experiments can easily be indexed using bcc crystal structure (space group $Im\bar{3}m$). Miller indices are assigned to all observed XRD peaks. The values of refinement parameters are generally used to estimate the quality of the fit ($\chi^2$ = 3.5763, $R_{wp}$ = 3.73%, and $R_p$ = 2.82%), demonstrating the high quality of the obtained samples. The lattice parameters are obtained to be $a$ = 3.445(1) by analyzing the diffraction data. Figure 1(b) is a typical SEM image of a fresh cross-section for the TiHfNbTaMo HEA sample. A single-color back-scattered electron image indicates no impurity phases (see Figure 1(c)). The ratio of Ti/Hf/Nb/Ta/Mo is 0.90:0.85:1.06:1.14:1.05, which agrees with the starting composition. The slight deviations are attribute to the unevenness of the cross-section, the overlap of EDX peaks in the emission spectra, and the limited accuracy of EDX. A microscopically homogeneous mixture of five elements is observed in TiHfNbTaMo HEA based on the elemental mappings.

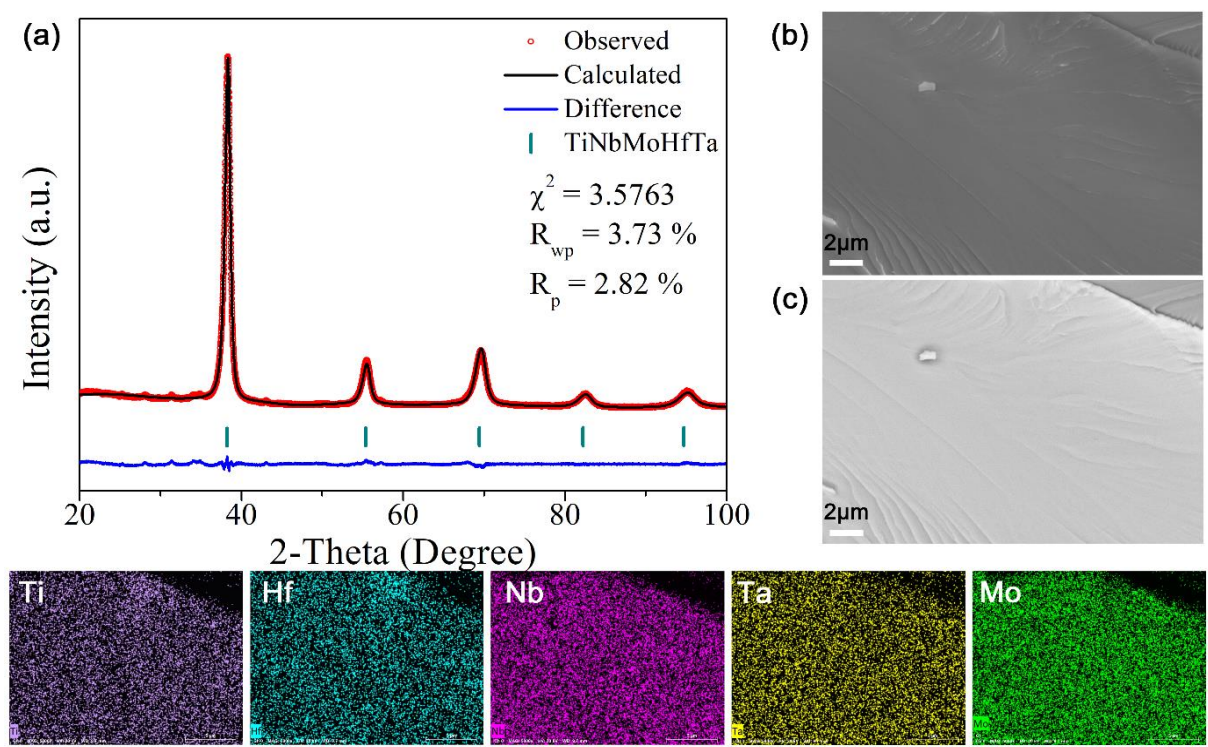

**Figure 1.** a) Rietveld refinement profile of the PXRD for TiHfNbTaMo HEA. b) SEM, c) BSEM, and EDX elemental mappings of the TiHfNbTaMo HEA.

The superconductivity in the TiHfNbTaMo HEA is investigated through magnetic measurements (see Figure 2(a)-(d)). Fig. 2(a) exhibits the temperature-dependent volume dc magnetic susceptibility $\chi_v(T)$ under the field-cooled (FC) and zero-field-cooled (ZFC) process. An obvious transition to the superconducting state is observed in the ZFC measurement. A pronounced diamagnetic signal is observed below the superconducting transition temperature $T_c$ = 3.42 K. When the diamagnetic factor (discussed subsequently) is considered, the ZFC measurements agree with a 100 % shielding fraction. Figure 2(b) illustrates the volume magnetization versus applied magnetic field curves at 1.8 - 3.5 K. With the assumption that a perfect diamagnet responds linearly to a magnetic field, a linear function has been fitted to the M(H) data at the lowest temperature (1.8 K): $M_{fit} = eH + f$. The demagnetization factor ($N$) is determined by solving the formula $-e = \frac{1}{4\pi(1-N)}$, resulting in a value of $N$ = 0.20. The demagnetization factor accounts for a sample shape-dependent magnetic field distortion inside and around a sample. Figure 2(c) presents the M-$M_{fit}$ curves at various temperatures. Lower critical fields $\mu_0H_{c1}*$ are considered deviations in the linearity of magnetic field response in the respective fields. The $\mu_0H_{c1}*$ values derived for various temperatures are plotted versus temperature in Figure 2(d) and modeled with the formula $\mu_0H_{c1}^*(T) = \mu_0H_{c1}^*(0)(1-(T/T_c)^2)$. The critical field extrapolated to T = 0 K through the formula gives $\mu_0H_{c1}^*(0)$ = 18.2 mT. Considering the demagnetization factor $N$, the lower critical field at 0 K was estimated using this expression $\mu_0H_{c1}(0) = \mu_0H_{c1}^*(0)/(1-N)$. The obtained value $\mu_0H_{c1}(0)$ is 22.8 mT for TiHfNbTaMo HEA.

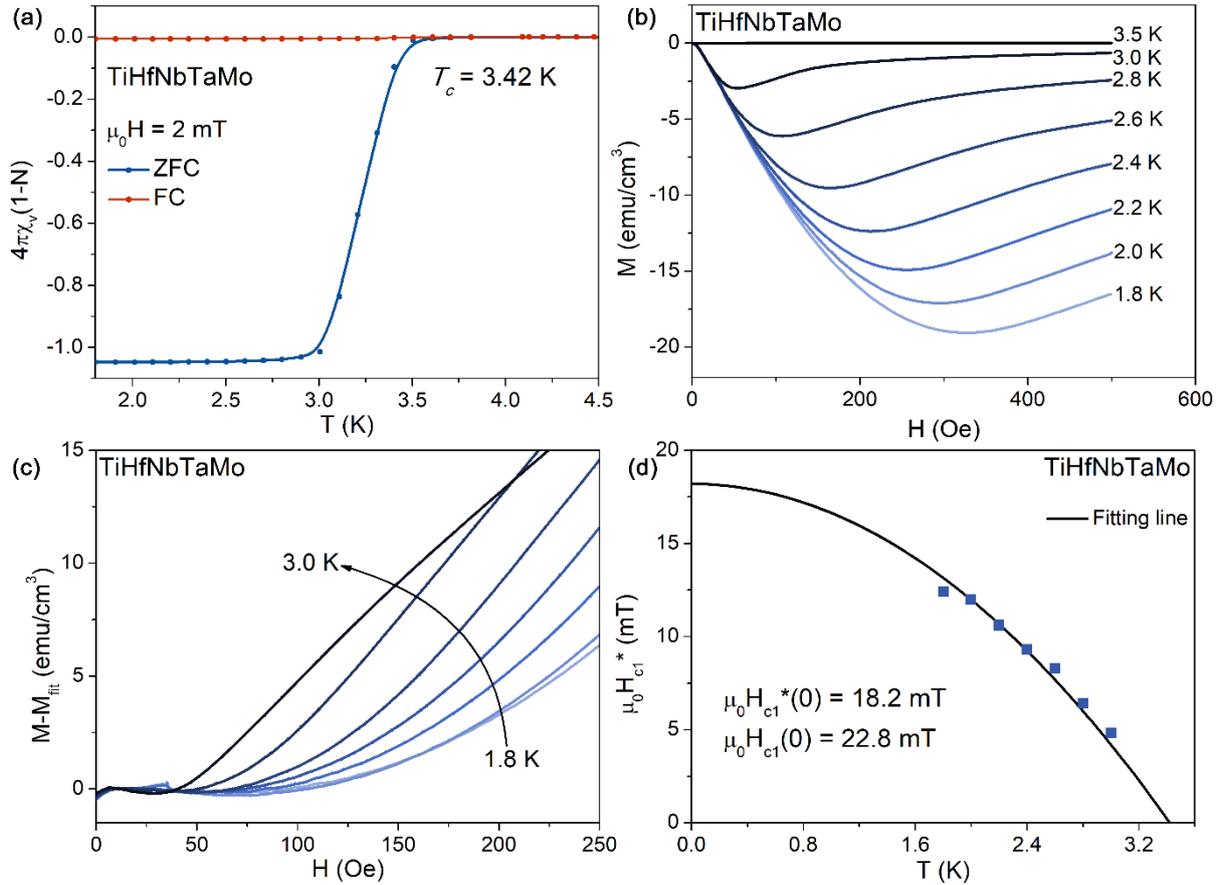

**Figure 2.** Superconducting properties of TiHfNbTaMo HEA a) Temperature-dependent magnetization curves. b) M(H) curves at a temperature ranging from 1.8 to 3.5 K with external

fields varying from 0 to 500 Oe. c) The variation of M and $M_{fit}$ from 0-250 Oe at 1.8 - 3.0 K. d) The temperature dependence of the $\mu_0 H_{c1}^*$.

Figure 3(a) shows the temperature-dependent electrical resistivity behavior for the TiHfNbTa HEA over a temperature between 1.8 - 300 K, with measurements in a zero-applied magnetic field. The resistivity in the normal state exhibits a weak metallic temperature dependence, which can be ascribed to the existence of atomic disorders. Additionally, the relatively low residual resistivity ratio RRR of approximately 1.18 is due to the HEA sample's considerable structural disorder and polycrystalline nature. At lower temperatures, a distinct superconducting transition is observed at around 3.87 K, with zero resistivity occurring at 3.57 K, slightly surpassing the $T_c$ value of 3.42 K obtained from the magnetic measurement. When a magnetic field is applied, the $T_c$ moves toward lower temperatures, as depicted in the inset of Figure 3(a). The midpoints of these transitions, obtained from the resistivity data, were utilized to estimate the upper critical field, as illustrated in Figure 3(b). We can still observe superconductivity onset transition under a 3 T applied magnetic field. Figure 3(b) shows the temperature-dependent upper critical fields $\mu_0 H_{c2}(T)$. It can be observed that the upper critical field $\mu_0 H_{c2}$ exhibits a nearly linear relationship with decreasing temperature. The corresponding slope, represented as $d\mu_0 H_{c2}/dT$, is determined to be -1.252 T/K. Werthamer-Helfand-Hohenberg (WHH) formula was used to fit the data: $\mu_0 H_{c2}(0) = -0.693 T_c (\frac{d\mu_0 H_{c2}}{dT})|_{T=T_c}$. The dirty limit $\mu_0 H_{c2}(0)^{WHH}$ value was found to be 3.29 T.

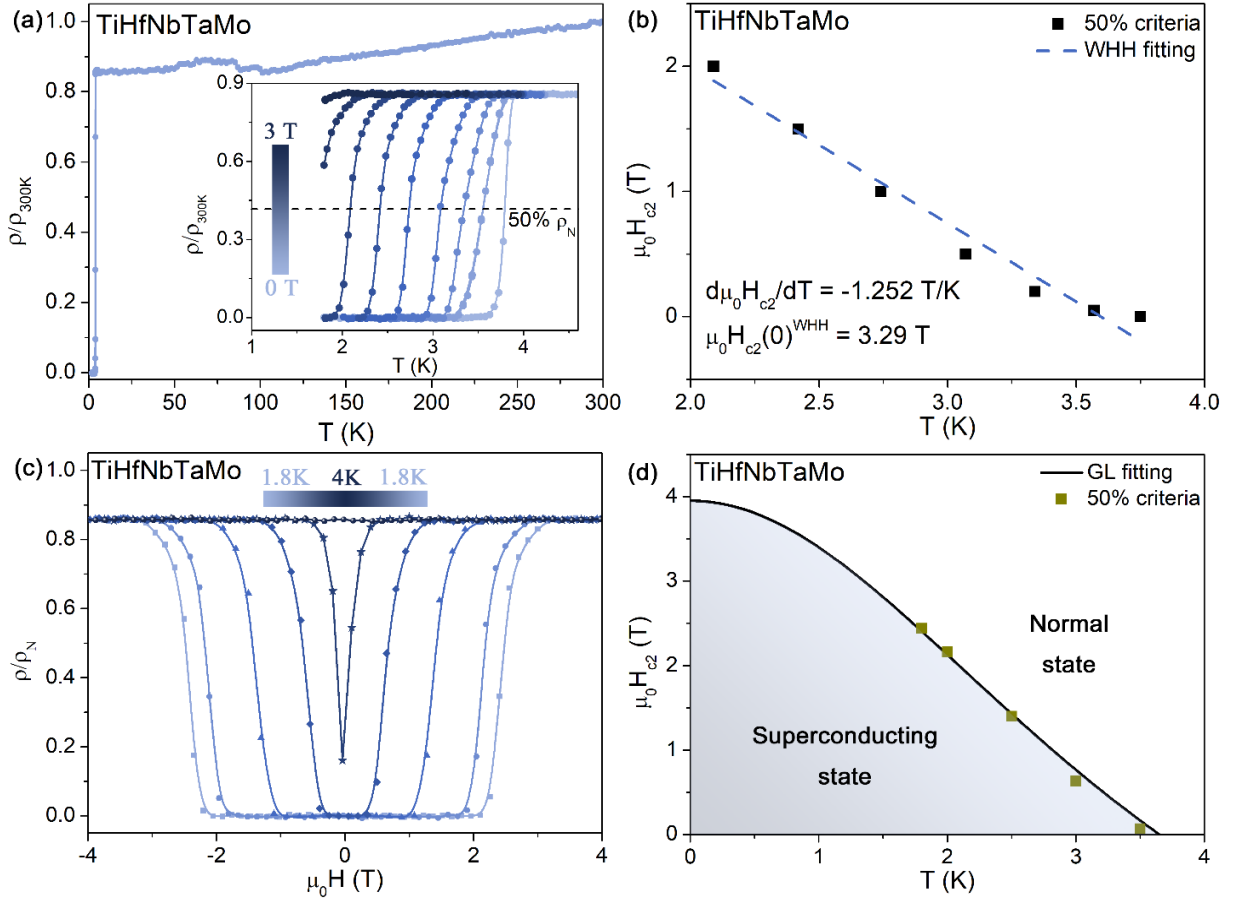

**Figure 3.** The electric transport performance of TiHfNbTaMo HEA. a) Electrical resistivity at zero applied magnetic fields. An inset shows the field-dependent resistivity between 0 and 3 T near the transition to superconductivity. b) The temperature dependence of the $\mu_0H_{c2}$ and fitting with the WHH model. c) Perpendicular field magnetoresistance at different temperatures in fields between -4 and 4 T. d) The temperature-dependent $\mu_0H_{c2}$ was fitting with the GL model.

We also measured magnetotransport under a perpendicular field at various temperatures to explore the superconductivity nature of the TiHfNbTaMo HEA (see Figure 3(c)). The $\mu_0H_{c2}$ corresponds to the resistivity value down to 50 % of $\rho_N$. Fig. 3(d) presents the phase diagram of T versus $\mu_0H_{c2}$. Ginzburg-Landan (GL) formula was used to fit the data: $\mu_0H_{c1}^*(T) = \mu_0H_{c1}^*(0)(1-(T/T_c)^2)$. The GL model satisfactorily fits experimental data over the entire temperature range. The $\mu_0H_{c2}(0)^{GL}$ can be written as 3.95 T. Both WHH and GL $\mu_0H_{c2}(0)$ are lower than, and therefore in agreement with the Pauli paramagnetic limit, where $\mu_0H^P = 1.85*T_c$ = 6.33 T for the TiHfNbTaMo HEA sample.

Based on the results of $\mu_0H_{c1}(0)$ = 22.8 mT and $\mu_0H_{c2}(0)$ = 3.95 T, we can calculate and extract various superconducting parameters for the TiHfNbTaMo HEA. The GL coherence length, $\xi_{GL}(0)$, can be derived via the expression $\xi_{GL}^2(0) = \frac{\Phi_0}{2\pi\mu_0H_{c2}(0)}$, where the quantum flux $\phi_0 = \frac{h}{2e} = 2.07 \times 10^{-15}$ Wb. The $\xi_{GL}(0)$ is calculated to be 91.33 Å, which is close to that seen in $Mo_{0.11}W_{0.11}V_{0.11}Re_{0.34}B_{0.33}$ (84 Å)[42] and $Cr_{14}Mo_{26}W_{12}Re_{35}Ru_{13}C_{20}$ (79 Å)[24] HEA superconductors. Subsequently, the GL penetration depth at 0 K, $\lambda_{GL}(0)$, is estimated as 1405 Å using the relation $\mu_0 H_{c1}(0) = \frac{\Phi_0}{4\pi\lambda_{GL}^2(0)} ln\frac{\lambda_{GL}(0)}{\xi_{GL}(0)}$. Accordingly, from the formula $K_{GL}(0) = \frac{\lambda_{GL}(0)}{\xi_{GL}(0)}$, we obtain the GL parameter $K_{GL}(0) = 15.4$. The value is greater than $1/\sqrt{2}$, confirming the TiHfNbTaMo HEA is a type-II superconductor. The thermodynamic critical field ($\mu_0 H_c(0)$) is determined from $\mu_0 H_{c1}(0) \times \mu_0 H_{c2}(0) = \mu_0 H_c^2(0) \times lnK_{GL}(0)$ as 0.18 T.

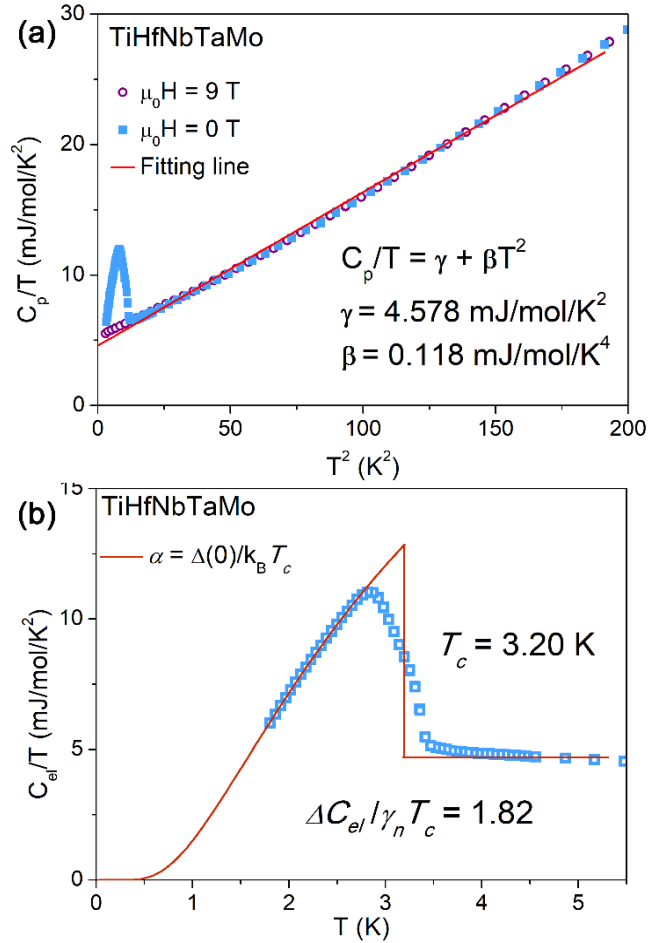

**Figure 4.** Specific heat curves under 0 and 9 T magnetic field of TiHfNbTaMo HEA. a) Specific heat coefficient $C_p/T$ as a function of $T^2$. b) electronic contribution to the heat capacity as $C_{el}/T$.

Further support for the bulk superconducting state in TiHfNbTaMo HEA was obtained from the heat capacity measurements under the applied magnetic field of 0 and 9 T, as presented in Figure 4(a). The experimental point was fitted in the normal state (above $T_c$) to the formula $C_p/T = \gamma_n + \beta T^2$, where $\gamma_n$ represents the Sommerfeld constant of the normal state, and $\beta$ is the specific heat coefficient of the lattice part. The red line in Figure 4(a) shows the data fitted by this formula and yields estimates of $\beta$ = 0.118 mJ/mol/K$^4$ and $\gamma_n$ = 4.578 mJ/mol/K$^2$. The specific heat jump was suppressed completely under the magnetic field of 9 T, indicating that $\mu_0 H$ = 9 T is higher than the upper critical field. The Debye model was then used with the value in the formula $\Theta_D = (12\pi^4 nR/5\beta)^{1/3}$ to estimate the Debye temperature $\Theta_D$, where $R$ represents the gas constant and $n$ signifies the number of atoms in the formula unit. The $\Theta_D$ was estimated to be 254 K.

The temperature-dependent electronic specific heat ($C_{el}$) is shown in Figure 4(b). The so-called α-model is employed to analyze the data. The α-model still assumes a fully isotropic superconducting gap but allows for the variation of coupling constant $\alpha = \Delta(0)/k_B T_c$, where $\Delta(0)$ represents the gap size at 0 K. Considering the entropy conservation at $T_c$, bulk $T_c$ = 3.20 K under a zero applied field is determined, which aligns with the $T_c$ value obtained from resistivity and magnetic susceptibility data. With the entropy-conserving construction displayed in Figure 4(b), the normalized specific heat jumps at $T_c$, $\Delta C_{el}/\gamma_n T_c \sim 1.82$, is evidencing larger than the BCS value of 1.43 and closely approximates the value of Nb element (1.87).[43] Consequently, TiHfNbTaMo can be classified as moderately strong coupling superconductors for Cooper pairing. Within the α-model, the angular independent gap function $\Delta(T)$ is expressed to be $\Delta(T) = \alpha/\alpha_{BCS}\Delta_{BCS}(T)$, where $\alpha_{BCS}$ = 1.76 represents the weak coupling gap ratio. The fitting of the $C_{el}$ data, as shown in Figure 5(b), yields a superconducting gap value $\Delta(0)$ = 0.55 meV at zero temperature. The calculation of the coupling strength $2\Delta_0/k_B T_c$ yields a value of 3.98, surpassing the BCS value of 3.52 for weak coupling. Furthermore, the estimation of the electron-phonon coupling constant $\lambda_{ep}$ for TiHfNbTaMo HEA is derived from the $\Theta_D$ value, employing the semiempirical McMillan formula $\lambda_{ep} = \dfrac{1.04 + \mu^* \ln\left(\dfrac{\Theta_D}{1.45T_c}\right)}{(1 - 0.62\mu^*)\ln\left(\dfrac{\Theta_D}{1.45T_c}\right) - 1.04}$. The Coulomb pseudopotential parameter $\mu^*$ = 0.13,[44-46] is valid for intermetallic superconductors. Based on the $T_c$ and $\Theta_D$, we obtained $\lambda_{ep}$ = 0.60 for TiHfNbTaMo HEA, which allows us to classify the TiHfNbTaMo HEA as a superconductor exhibiting moderate electron-phonon coupling.

The electronic band structure of TiHfNbTaMo shows a good metallic feature with two bands crossing the Fermi level, see Figure 5(a), in agreement with the experimental measurements. In addition, some saddle-point-like bands exist around the Fermi level, such as that at ~$E_F$-0.7 eV near point P, ~$E_F$-0.1 eV, and ~$E_F$+0.2 eV near point N in the energy dispersion. Such band structure usually corresponds to van Hove singularity and a large peak in the density of states (DOSs). We plot the total DOSs and orbital projected DOSs in Figure 5(b) to check this phenomenon. One can see that the DOSs are dominated by hybridized $d$-orbitals from these five elements, with a DOS of about 2.18 states eV$^{-1}$ f.u.$^{-1}$.

Moreover, three van Hove singularities exist, as labeled by the red dots in Figure 5(b). Substitute the DOS into the relationship $1/3\pi k_B N_A N(E_F)$, the theoretical Sommerfeld coefficient $\gamma$ can be estimated as 5.14 mJ mol$^{-1}$ K$^{-2}$, which is slightly larger with the experimental value 4.578 mJ mol$^{-1}$ K$^{-2}$. Since the van Hove singularity at the Fermi level is believed to be an essential reason leading to the system's instability along with phase transitions, TiHfNbTaMo has a strong possibility of superconductivity when temperature decreases to a critical point.

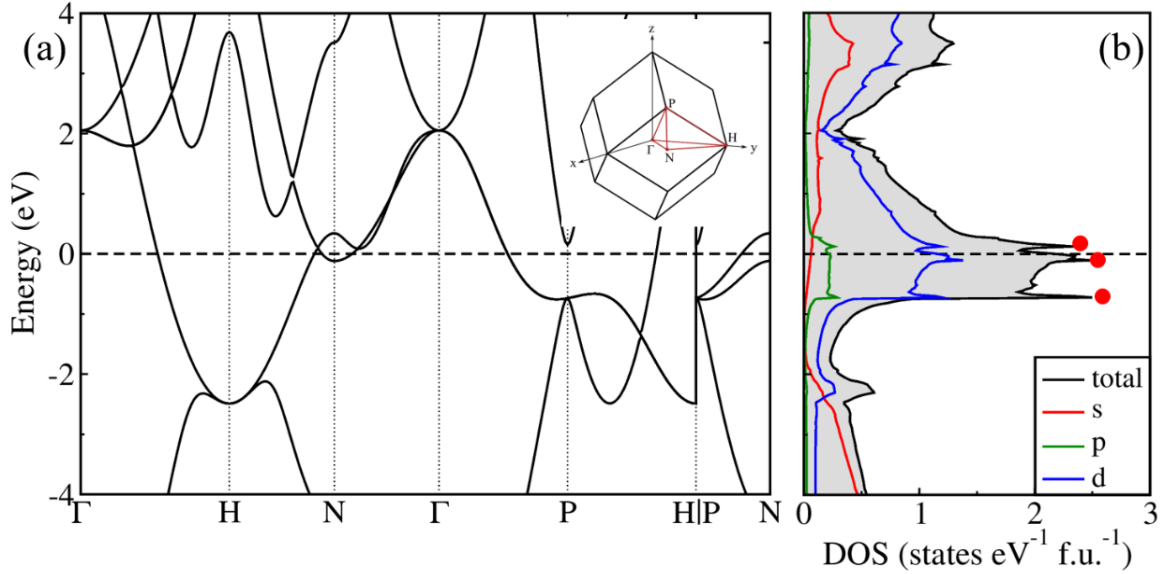

**Figure 5.** a) Calculated electronic structure of TiHfNbTaMo HEA. b) Total DOS and orbital projected DOS of TiHfNbTaMo HEA.

Figure 6 shows the $T_c$ with the VEC of the bcc-type HEA superconductors.[2,3,14-22] For comparison, the observed trend lines of the $T_c$ for crystalline 4$d$ metals (blue dotted line) and amorphous 4$d$ metals (green dotted line) are also depicted.[39,47] The trend of transition metals is commonly known as the Matthias rule, where the highest $T_c$ is achieved at a VEC range of 4.5-4.7. As displayed in Figure 6, the $T_c$ with VEC of the TiHfNbTaMo is comparable to the other reported bcc-type HEA superconductors. In some bcc-type HEA systems, the $T_c$ and VEC relationship roughly follows the behavior of the 4$d$ crystalline metallic superconductors, such as (TaNb)$_{1-x}$(HfZrTi)$_x$ system and (TaNb)$_{0.67}$(HfZrTi)$_{0.33}$Al$x$ system.[16,17] Although the VEC for most HEAs with a $T_c$ higher than 7 K is around 4.67, the relationship between $T_c$ and VEC cannot be described by Mattias's rule for different HEA systems, which have different compositions of elements. As can be seen from Figure 6, the range of $T_c$ is distributed in 2.8-7.8 K when VEC = 4.67. Table S1 summarizes the parameters of HEA for all single-phase bcc structures with $T_c$ higher than 7 K. It can be seen that the $T_c$ is significantly affected by the elemental makeup. Moreover, elemental Nb has been observed to play a significant role in determining the $T_c$ of bcc-type HEA superconductors. In the Ti-Zr-Hf-Nb-Ta system, where isoelectronic substitutions are made, a direct correlation has been established between the Nb content and $T_c$, with higher Nb content resulting in higher $T_c$.[16] The presence of Nb substantially enhances $T_c$ in bcc-type HEA superconductors at the same VEC. When the $T_c$ of bcc-type HEA is higher than 7 K, the content of the Nb element is higher than or equal to 25%. This result seems predictable since Nb is all elements' highest $T_c$ (~9.2 K). Hence, the VEC and elemental composition influence the bcc-type HEA's $T_c$, particularly the elemental Nb content. A VEC approximately equal to 4.6 and a substantial proportion of Nb elements are conducive to elevating the $T_c$ of the bcc-type HEA. TiHfNbTaMo has a VEC of 4.8 and an Nb content of 20 %. Thus, its $T_c$ is only 3.42 K.

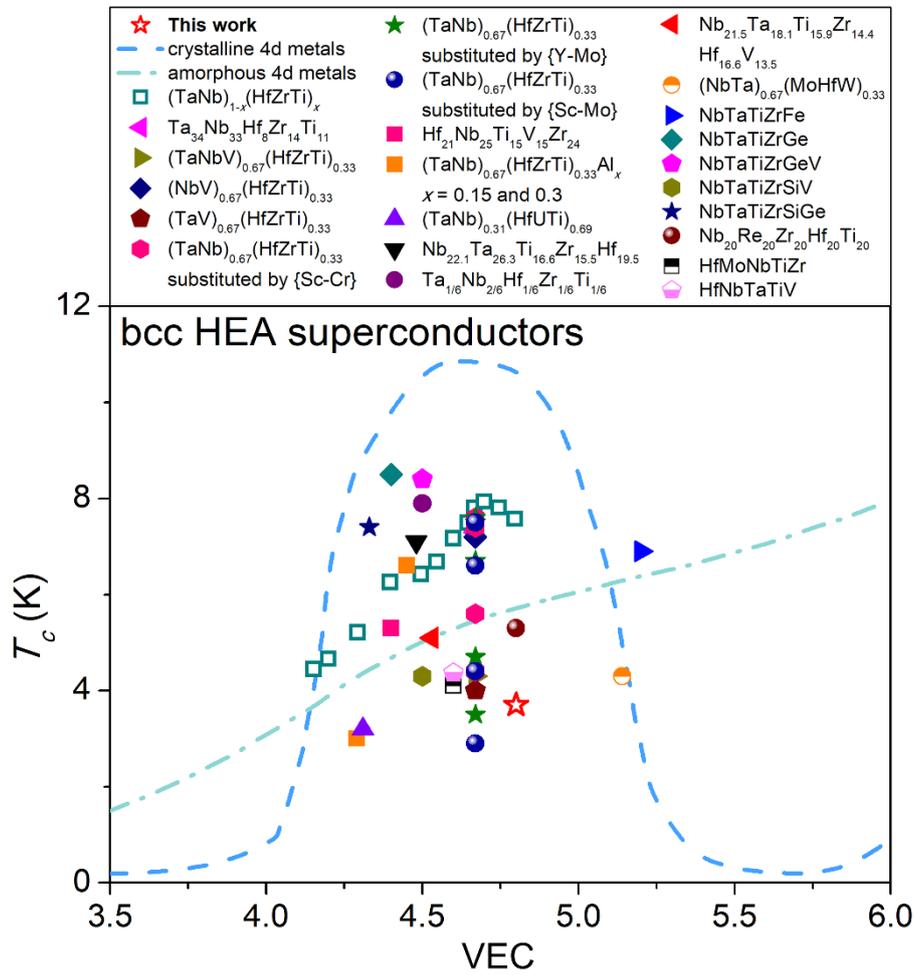

**Figure 6.** Critical temperature $T_c$ with VEC of the bcc-type HEA superconductors, [2,3,14-22] crystalline 4d metals (blue dotted line), and amorphous 4d metals (green dotted line). [39,47]

Despite the presence of highly disordered atoms on simple lattices in bcc-type HEA superconductors, the superconducting properties of these materials are significantly influenced by their elemental composition and VEC. Published evidence focusing on the electronic properties of these materials suggests that similar to other HEA counterparts, HEA superconductors possess exceptional mechanical properties that could be valuable for employment under harsh environments. Meanwhile, though the upper critical fields of the current HEA superconductors are not as high as those of $Nb_3Sn$ or NbTi, which are currently utilized in the production of the majority of commercial superconducting magnets, we anticipate that future superconducting HEAs could serve as promising materials for the fabrication of such magnets. Furthermore, we hold the belief that this field has boundless potential for the exploration of novel phenomena.

## 4. Conclusion

In conclusion, the formation and superconducting properties of a HEA, namely TiHfNbTaMo, with a VEC of 4.8, are described. The crystal structure refinement presented that the alloy forms a simple bcc structure, as anticipated for HEA, and a microprobe analysis revealed a uniform distribution of the five metal elements. This material shows type-II superconductivity with $T_c$ = 3.42 K, upper critical field $\mu_0H_{c2}(0)$ = 3.95 T, and lower critical field $\mu_0H_{c1}(0)$ = 22.8 mT. Low-temperature specific heat measurements suggest that the alloy is a conventional s-wave type with a moderately coupled superconductor. First-principles calculations show that the DOS of the TiHfNbTaMo alloy is dominated by hybrid $d$ orbitals of these five metal elements. Additionally, the TiHfNbTaMo HEA exhibits three van Hove singularities. Finally, it found that the VEC and the composition of the elements (especially the Nb elemental content) together affect the $T_c$ of bcc-type HEA.

**Acknowledgments**
This work is supported by the National Natural Science Foundation of China (Grants No. 11922415, 11974432, 52271016, 52188101), Guangdong Basic and Applied Basic Research Foundation (2022A1515011168, 2019A1515011718). M. Boubeche is supported by China's Foreign Young Talents Program (22KW041C211). K. Jin is supported by the Key-Area Research and Development Program of Guangdong Province (Grant No. 2020B0101340002).

H. Luo, *Sci. China Phys. Mech. Astro*. **2023**. *66*, 277412.

[47] M. M. Collver, R. H. Hammond, *Phys. Rev. Lett*. **1973**, *30*, 92.

[48] G. Kresse, J. Furthmüller, *Phys. Rev. B* **1996**, *54*, 11169.

[49] J. P. Perdew, K. Burke, M. Ernzerhof, *Phys. Rev. Lett.* **1997**, *77*, 3865.

[50] L. Bellaiche, D. Vanderbilt, *Phys. Rev. B* **2000**, *61*, 7877.

# Supporting Information

# Superconductivity in the bcc-type High-entropy Alloy TiHfNbTaMo

*Lingyong Zeng[1#], Jie Zhan[2,3 #], Mebrouka Boubeche[4], Kuan Li[1], Longfu Li[1], Peifeng Yu[1], Kangwang Wang[1], Chao Zhang[1], Kui Jin[4,5,6], Yan Sun[2, 3*], Huixia Luo[1,*]*

[1]School of Materials Science and Engineering, State Key Laboratory of Optoelectronic Materials and Technologies, Key Lab of Polymer Composite & Functional Materials, Guangdong Provincial Key Laboratory of Magnetoelectric Physics and Devices, Sun Yat-Sen University, No. 135, Xingang Xi Road, Guangzhou, 510275, P. R. China

[2]Shenyang National Laboratory for Materials Science, Institute of Metal Research, Chinese Academy of Sciences, Shenyang, 110016, China;

[3]School of Materials Science and Engineering, University of Science and Technology of China, Shenyang 110016, China

[4]Songshan Lake Materials Laboratory, Building A1, University Innovation Town, Dongguan City, Guang Dong Province, 523808 China

[5]Beijing National Laboratory for Condensed Matter Physics, Institute of Physics, Chinese Academy of Sciences, Beijing, China

[6]Key Laboratory of Vacuum Physics, School of Physical Sciences, University of Chinese Academy of Sciences, Beijing, China

([#]L. Zeng and J. Zhan contributed equally to this work)
E-mail: luohx7@mail.sysu.edu.cn; or sunyan@imr.ac.cn

**Table S1** the $T_c$, VEC, and Nb content of the bcc-type HEAs with $T_c$ higher than 7 K.

| bcc-type HEA | $T_c$ | VEC | Nb content |
| --- | --- | --- | --- |
| $Ta_{34}Nb_{33}Hf_8Zr_{14}Ti_{11}$[1] | 7.30 K | 4.67 | 33.0% |
| $(TaNb)_{0.6}(HfZrTi)_{0.4}$[2] | 7.17 K | 4.60 | 25.0% |
| $(TaNb)_{0.65}(HfZrTi)_{0.35}$[2] | 7.49 K | 4.65 | 27.7% |
| $(TaNb)_{0.67}(HfZrTi)_{0.33}$[2] | 7.75 K | 4.67 | 28.7% |
| $(TaNb)_{0.7}(HfZrTi)_{0.3}$[2] | 8.03 K | 4.70 | 30.0% |
| $(TaNb)_{0.75}(HfZrTi)_{0.25}$[2] | 7.81 K | 4.74 | 33.3% |
| $(TaNb)_{0.8}(HfZrTi)_{0.2}$[2] | 7.57 K | 4.79 | 36.4% |
| $(TaV)_{0.67}(HfZrTi)_{0.33}$[2] | 7.20 K | 4.67 | 28.7% |
| $(TaNb)_{0.67}(Sc_{0.67}Cr_{0.33}ZrTi)_{0.33}$[3] | 7.50 K | 4.67 | 28.7% |
| $(TaNb)_{0.67}(HfSc_{0.67}Cr_{0.33}Ti)_{0.33}$[3] | 7.40 K | 4.67 | 28.7% |
| $(TaNb)_{0.67}(HfZrSc_{0.67}Cr_{0.33})_{0.33}$[3] | 7.60 K | 4.67 | 28.7% |
| $(TaNb)_{0.67}(Y_{0.67}Mo_{0.33}ZrTi)_{0.33}$[3] | 7.60 K | 4.67 | 28.7% |
| $(TaNb)_{0.67}(Y_{0.67}Mo_{0.33}ZrHf)_{0.33}$[3] | 7.50 K | 4.67 | 28.7% |
| $(TaNb)_{0.67}(Sc_{0.67}Mo_{0.33}ZrTi)_{0.33}$[3] | 7.50 K | 4.67 | 28.7% |
| $(TaNb)_{0.67}(Sc_{0.67}Mo_{0.33}ZrHf)_{0.33}$[3] | 7.50 K | 4.67 | 28.7% |
| $Ta_{1/6}Nb_{2/6}Hf_{1/6}Zr_{1/6}Ti_{1/6}$[4,5] | 7.85 K | 4.92 | 33.0% |